\documentclass{article}
\usepackage{spconf,amsmath,graphicx}
\usepackage{multirow}

\usepackage{booktabs}
\title{decoupling pronunciation and language for end-to-end code-switching automatic speech recognition }
\name{Shuai Zhang$^{1,2}$, Jiangyan Yi$^{2}$, Zhengkun Tian$^{1,2}$, Ye Bai$^{1,2}$, Jianhua Tao$^{1,2,3}$, Zhengqi wen$^{2}$}
\address{
	$^1$School of Artificial Intelligence, University of Chinese Academy of Sciences, China \\
	$^2$NLPR, Institute of Automation, Chinese Academy of Sciences, China \\
	$^3$CAS Center for Excellence in Brain Science and Intelligence Technology, China\\}
\begin{document}
\ninept
\maketitle
\begin{abstract}
Despite the recent significant advances witnessed in end-to-end (E2E) ASR system for code-switching, hunger for audio-text paired data limits the further improvement of the models' performance. In this paper, we propose a decoupled transformer model to use  monolingual paired data and unpaired text data to alleviate the problem of code-switching data shortage. The model is decoupled into two parts: audio-to-phoneme (A2P) network and phoneme-to-text (P2T) network. The A2P network can learn acoustic pattern scenarios using large-scale monolingual paired data. Meanwhile, it generates multiple phoneme sequence candidates for single audio data in real time during the training process. Then the generated phoneme-text paired data is used to train the P2T network. This network can be pre-trained with large amounts of external unpaired text data. By using monolingual data and unpaired text data, the decoupled transformer model reduces the high dependency on code-switching paired training data of E2E model to a certain extent. Finally, the two networks are optimized jointly through attention fusion. We evaluate the proposed method on the public Mandarin-English code-switching dataset. Compared with our transformer baseline, the proposed method achieves 18.14\% relative mix error rate reduction.
\end{abstract}
\section{Introduction}
\label{sec:intro}
Code-switching is the phenomenon where speakers alternate between different languages within a sentence. With the progress of globalization, it is becoming an increasingly common linguistic behavior \cite{muysken2000bilingual}. For the code-switching ASR task, end-to-end (E2E) model is an emerging field because of its simplicity compared with the traditional ASR system \cite{li2019towards,8682674,8462201,DBLP:conf/nips/VaswaniSPUJGKP17}. E2E method integrates acoustic, pronunciation, and language models into a whole with joint optimization \cite{graves2006connectionist,graves2013speech,chan2016listen,8462506}. Despite the recent significant advances witnessed in E2E ASR system for code-switching, hunger for audio-text paired data limits the further improvement of the models' performance. 

To alleviate the problem of code-switching data shortage, monolingual data and data augment technology are usually used \cite{DBLP:journals/corr/abs-2006-00782,DBLP:conf/interspeech/KhassanovXPZCNM19,DBLP:conf/interspeech/YilmazHL18,DBLP:conf/interspeech/ChangCL19,DBLP:conf/sltu/YilmazH18}. For the monolingual, some E2E models have achieved great performance with large-scale data. However, these models usually can not handle the code-switching speech well. One potential reason is that the output of the model's decoder depends on the previous outputs. When the previous steps keep emitting tokens from one language, it is hard to switch to tokens of another language immediately due to this dependency \cite{li2019towards}. Therefore, it is often necessary to design a transfer learning method ingeniously to transfer useful information from monolingual data to the ASR model \cite{DBLP:journals/corr/abs-2006-00782,DBLP:conf/interspeech/KhassanovXPZCNM19}.  For the data augment technology, the main research work focuses on the code-switching text generation issue \cite{DBLP:conf/interspeech/YilmazHL18,DBLP:conf/interspeech/ChangCL19}. However, the unpaired text data can not directly be used to train E2E ASR model. So some methods like language model fusion or knowledge distillation are utilized to indirectly assist training and decoding \cite{DBLP:conf/interspeech/SriramJSC18,DBLP:conf/interspeech/ZhaoSRRBLP19,DBLP:conf/interspeech/BaiYTTW19}. To obtain audio-text paired data, speech synthesis technology is necessary \cite{DBLP:journals/corr/abs-2006-00782}. However, the synthesized speech usually does not match the real speech, which is a challenge to improve the code-switching ASR performance.

In this paper, we propose a decoupled transformer model to use  monolingual paired data and unpaired text data to alleviate the problem of code-switching data shortage. We get inspiration from the work \cite{DBLP:conf/icassp/ChenLLY18} and make further improvements more suitable for code-switching ASR tasks. Compared with the conventional transformer model, our model is decoupled into two parts: audio-to-phoneme (A2P) network and phoneme-to-text (P2T) network. The A2P network is pre-trained using the connectionist temporal classification (CTC) criterion. Because the CTC model has the output independence assumption, which may make it more desirable to model acoustic pattern scenarios as the current step output does not explicitly rely on previous outputs \cite{li2019towards}. So the A2P network can learn acoustic pattern scenarios utilizing large-scale monolingual data, eliminating the influence of contextual text. Meanwhile, it generates multiple phoneme sequence candidates for single audio example in real-time during the training process. Then the generated phoneme-text paired data is used to train the P2T network. This network can also be pre-trained with large amounts of external unpaired text data. And code-switching text can be obtained relatively easily through various text generation techniques for code-switching. By utilizing monolingual paired data and unpaired text data, the decoupled transformer model reduces the high dependency on code-switching paired training data of the E2E model. Finally, the two networks are optimized jointly through attention fusion. The proposed model can not only effectively use monolingual paired data and unpaired text data, but also maintains the simplicity of the training and decoding process as an E2E model.

In this work, our main contributions are as follows. Firstly, a decoupled transformer model is proposed to use monolingual paired data and unpaired text data. Secondly, a multi-level attention mechanism is used to reduce computational cost. Finally, the method maintains the simplicity of the training and decoding process as an E2E model. The experimental results on ASRU 2019 Mandarin-English code-switching Challenge dataset show that our method has consistent improvement compared with the baseline. It is an effective strategy for code-switching ASR tasks.

\begin{figure*}[htb]

	\begin{minipage}[b]{1.0\linewidth}
		\centering
		\centerline{\includegraphics[width=16.0cm]{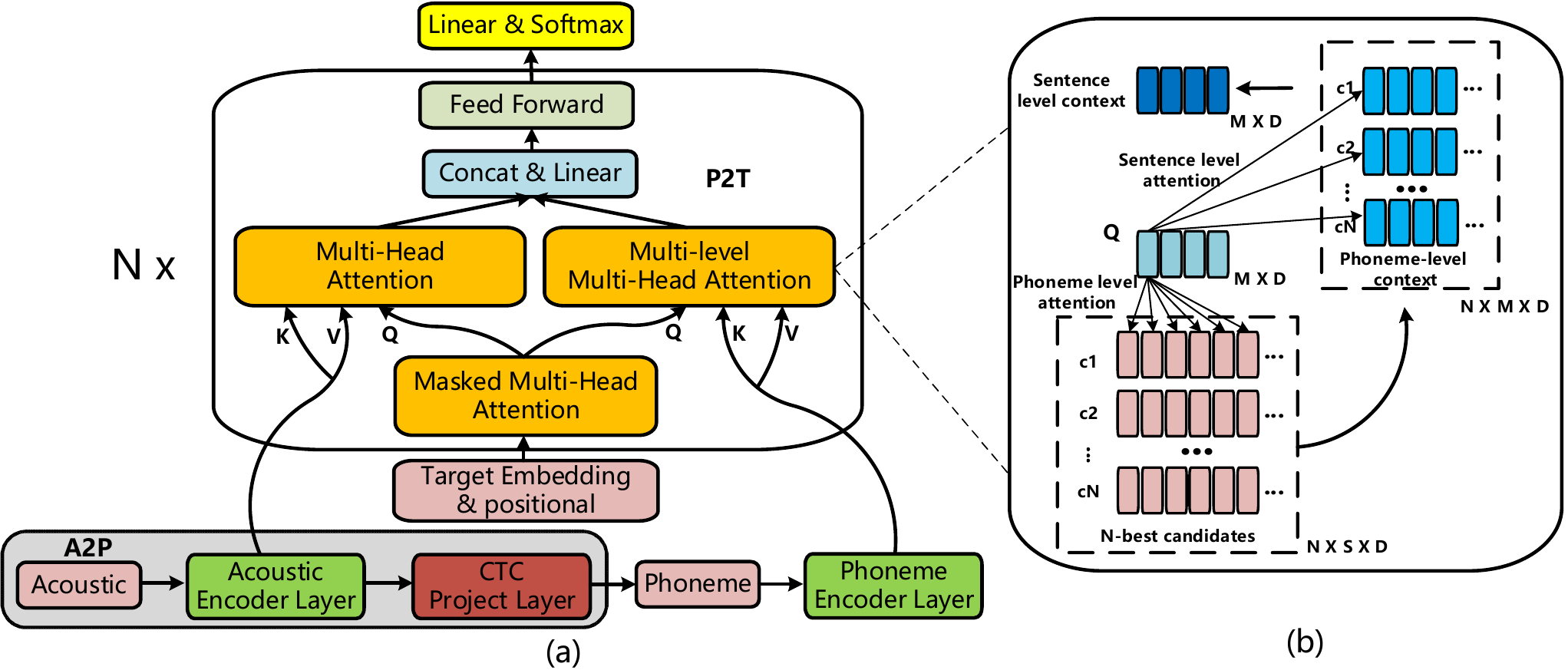}}
		%  \vspace{1.5cm}
		\centerline{}%\medskip
		\caption{Framework of decoupled transformer model (a) and multi-level attention (b).
			For simplicity, we omit parts such as convolution layer, layer normalization, residual connection, etc.}
	\end{minipage}

%	\begin{minipage}[b]{1.0\linewidth}
%	\centering
%	\centerline{\includegraphics[width=6.0cm]{figure-model8}}
%	%  \vspace{1.5cm}
%	\centerline{}%\medskip
%	\caption{Framework of decoupled transformer model (left) and multi-level attention (right).
%		For simplicity, we omit parts such as convolution layer, layer normalization, residual connection, etc.}
%	\end{minipage}
%	%	\hfill
	%	\begin{minipage}[b]{0.48\linewidth}
	%		\centering
	%		\centerline{\includegraphics[width=6.0cm]{figure-attention}}
	%		%  \vspace{1.5cm}
	%		\centerline{(b)}\medskip
	%	\end{minipage}
	
\end{figure*}

The rest of the paper is organized as follows. Section 2 reviews the the speech-transformer briefly. Section 3 describes the training strategy of decoupled transformer and multi-level attention in detail.  We briefly describe the related work and the differences between our method in Section 4. In Section 5, We present our experiments setups and results. Finally, we conclude our work in Section 6.

\section{Review of Speech-transformer}
\label{sec:pagestyle}
%In this section, we first introduce the basic structure of the speech-transformer model briefly. Then we introduce our model structure and training strategy in detail. Meanwhile, we describe the calculation process of the multi-level attention. 

Speech-transformer is a variant transformer model suitable for the ASR task \cite{8462506}. The details of its encoder and decoder are as follows.

For the encoder, 2D CNN layer is used to produce the acoustic hidden representation after down-sampling. After a linear layer, the positional encoding is used to attend relative positions. Then a stack of $N_{e}$ encoder-blocks is used to get the final encode representation. Each of the blocks has two sub-blocks: the first is multi-head attention whose queries, keys, and values come from the outputs of the previous block. And the second is position-wise feed-forward networks. Meanwhile, layer normalization and residual connection are introduced to each sub-block in training.

For the decoder, a learnable target embedding and positional encoding are applied to the target sequence. Then a stack of $N_{d}$ decoder-blocks is subsequent. Each decoder-block has three sub-blocks: the first is masked multi-head attention. And the masking is utilized to ensure the predictions depend only on the known outputs before positions. The second is a multi-head attention whose keys and values come from the encoder outputs and queries come from the previous sub-block outputs. Other parts are the same as encoder block. Finally, the outputs layer includes linear projection and soft-max layer.

\section{METHODS}
\label{sec:pagestyle}

\subsection{Decoupled Transformer}
\label{ssec:subhead}

The conventional E2E ASR method integrates acoustic, pronunciation, and language models into a whole with joint optimization. It directly maps the input sequence of acoustic features to the corresponding output sequence of targets. In this paper, we introduce intermediate phoneme sequences between acoustic input and target sequences. Then the model is decoupled into two parts: A2P network and P2T network. The benefit of this decoupling process is that the two networks can utilize large amounts of monolingual corpus and unpaired code-switching text data respectively. Therefore, this method reduces the high dependency on code-switching paired training data of the E2E model. The decoupling process can be expressed by the following formula.
\begin{equation}
P(\mathbf{x} | \mathbf{w}) \approx \max \limits_\mathbf{p} [P(\mathbf{w} | \mathbf{p},\mathbf{x}) \cdot P(\mathbf{p} | \mathbf{x})]
\end{equation}
where $\mathbf{w}$, $\mathbf{p}$, and $\mathbf{x}$ are word sequence, phoneme sequence and acoustic feature sequence respectively.  

For the training strategy, we first pre-train the A2P network using the CTC criterion. The modeling units are phoneme set of both Chinese and English. The A2P network is shown at the bottom left of Fig. 1(a). Then the A2P network is frozen as the acoustic encoder module of the whole model. In the training progress, the input acoustic features are firstly recognized to be multiple phoneme sequence candidates by the A2P model. These phoneme sequences and the target sequences form new phoneme-text paired training data. Then the generated phoneme-text data is used to train the P2T network. Besides, the P2T network can also be pre-trained with large amounts of external unpaired code-switching text data. The right half of the Fig. 1(a) P2T model represents the structure of the P2T.

As shown in the Fig. 1(a), keys and values from acoustic encoder and phoneme encoder are transmitted to the decoder block at the same time. The left is the normal multi-head attention calculation process for the acoustic encoder. And the right is the multi-level multi-head attention. We will describe the mechanism in detail in the next subsection. Then the context vectors from the acoustic encoder and phoneme encoder are concatenated, and the subsequent linear layer transforms it into a suitable dimension. In this way, the model combines the pronunciation and language information from monolingual data and unpaired text data. 

\subsection{Multi-level Attention}
\label{ssec:subhead}

For a single audio-text paired example, the proposed method will generate many intermediate phoneme sequences in the training process, e.g. the beam-size of CTC is set to 20. The intuitive approach is to complete the entire forward calculation and model optimization for each phoneme sequence. This greatly increases the computational cost of model training. Therefore, we propose a multi-level attention mechanism to reduce computational cost without compromising the performance. As shown in Fig. 1(b), phoneme level and sentence attentions are calculated sequentially. The $\mathbf{Q}$ is the queries of target encoding in Fig. 1(a). After the A2P decoding process and phoneme encoder layer, there are $\mathbf{N}$ phoneme sequence candidates. They are represented in Fig. 1(b) as an array of squares in the dashed box at the bottom left corner. For the phoneme level attention, the target queries $\mathbf{Q}$ calculates the attention weights with elements of each phoneme sequence, then $\mathbf{N}$ phoneme level context vector sequences are obtained. They are represented in Fig. 1(b) as an array of squares in the dashed box at the upper right corner. This process is the same as the usual attention calculation method. 

In order to fuse the information of $\mathbf{N}$ phoneme level context sequences into one sequence, an element-wise weighted average of different context sequences is used. And the weighting parameters can be learned by sentence level attention. Concretely, the first element of target query $\mathbf{Q}$ calculates the attention weights with the first elements of $\mathbf{N}$ phoneme level context sequences, then get the first element of final sentence level context after weighting. And repeat this process for all elements to complete the attention calculation.

\section{related work}
\label{sec:typestyle}

\subsection{Modular Trained End-to-End Model}
\label{ssec:subhead}
We get inspiration from the work \cite{DBLP:conf/icassp/ChenLLY18} and make further improvements more suitable for code-switching ASR tasks. There are three main differences between them.

1. The motivation of the decoupled transformer model is to use monolingual paired data and unpaired text data to alleviate the problem of code-switching data shortage. However, the work \cite{DBLP:conf/icassp/ChenLLY18} focus on modular-trained E2E framework, while E2E decoding is retained.

2. Our method unifies the A2P and P2T networks under the transformer framework. The P2T network fusion the acoustics information from A2P in training and decoding process. On the contrary, the acoustics representation A2P is not used by P2T model in the work \cite{DBLP:conf/icassp/ChenLLY18}. This may be one of the factors limiting the performance of the model.

3. Multi-level attention mechanism for word and sentence is proposed to reduce computational cost.

\section{EXPERIMENTS}
\label{sec:typestyle}
%In this section, we introduce in detail the data set and the model setups of our experiment. Then we present the experimental results and make discussion for them.

\subsection{Datasets}
\label{ssec:subhead}
We conduct our experiments on ASRU 2019 Mandarin-English code-switching Challenge dataset, which consists of about 200 hours code-switching training data and 500 hours Mandarin data \cite{DBLP:journals/corr/abs-2007-05916}. The development set and test set each has about 20 hours of data. All the data are collected in quiet rooms by various Android phones and iPhones. The transcripts of data cover many common fields including entertainment, travel, daily life and social interaction. Meanwhile, we choose the 460 hours of a subset of Librispeech corpus \cite{DBLP:conf/icassp/PanayotovCPK15} as the English data.

\subsection{Experiment Setups}
\label{ssec:subhead}

In this paper, the input acoustic features of the encoder network are 40-dimensional filter-bank with 25ms windowing and 10ms frame shift. For the output target, we adopt English word pieces and Chinese characters as the modeling units. We keep the 3277 characters that appear more than five times in the training set as the Chinese modeling units. And the number of English word pieces is 2k. The word pieces can not only balance the granularity of Chinese and English modeling units but also alleviate the out-of-vocabulary (OOV) problem with limited English training data. In this paper, we use a mix error rate (MER) to evaluate the experiment results of our methods. The MER is defined as the word error rate (WER) for English and character error rate (CER) for Mandarin.

We adopt the speech-transformer model as our baseline system. Two $3*3$ CNN layers with stride 2 for both time and frequency dimensions of the acoustic features are used. For both encoder-block and decoder-block, the attention dimension is 512 and the head number is 8. The dimension of position-wise feed-forward networks is 2048. And the number of encoder-block and decoder-block are 12 and 6 respectively. 

In order to eliminate the influence of model parameters' size, the blocks number of acoustic encoder and phoneme encoder are set to 8 and 4 in the decoupled transformer model. And the decoder-blocks number is 6. This setting ensures that the number of the proposed model parameters is close to the baseline. 

The A2P network is pre-trained using CTC criteria, and the output units are phonemes. The phonemes set has 208 Chinese phonemes and 84 English phonemes. Additionally bound separator $<wb>$ are used to indicate word boundaries in phoneme sequences. The motivation is to take $<wb>$ as the hint of tokenization in the phoneme-to-text process, e.g. distinguishing short words in case that its phoneme sequence is a sub-string of longer words \cite{DBLP:conf/icassp/ChenLLY18}. For the phoneme encoder network, a learnable phoneme embedding and positional encoding is applied, the dimension is 512. The rest of the phoneme encoder has the same structure as the acoustic encoder. Except for the multi-level attention mechanism described in section 3.2, the rest of the decoder-block also has the same structure as the decoder-block of baseline. 

The uniform label smoothing technique is used and the parameter is set to 0.1. SpecAugment is used to improve the performance of the models \cite{DBLP:conf/interspeech/ParkCZCZCL19}. Meanwhile, we set residual dropout as 0.1, where the residual dropout is applied to each sub-block before adding the residual information. We adopt the optimization strategy of work \cite{8462506} in the training process. After training, we average the last 5 checkpoints as the final model. Then, we performed decoding using beam search with a beam size of 10. All the experiments are conducted using ESPNet \cite{DBLP:journals/corr/abs-1804-00015}.

\subsection{Performance of A2P network}
\label{ssec:subhead}
The phoneme sequences generated by A2P network is used to train P2T part. So phoneme error rate (PER) of A2P model affects the final MER of the proposed method. Table 1 shows the results of the A2P model under different combinations of data sets. For the code-switching data, the A2P model achieves 5.45\% PER performance. And the Chinese training data improves the PER to 4.28\%. However, English data hurts the model performance slightly. The reason may be that the Chinese have accent problems in English words pronunciation. This leads to a mismatch problem between the English and code-switching data sets. The results show that the A2P model can model code-switching acoustic pattern scenarios effectively. It is the basis of the subsequent method.

\begin{table}[htb] 
	\centering 
	\caption{The phoneme error rate (PER \%) of pre-train CTC under different datasets. There are 200 hours code-switching data (200h CS), 500 hours Chinese data (500h CH), 460 hours English (460h EN) and all of them (All).} \label{tab:aStrangeTable} %
	\renewcommand\tabcolsep{9.5pt}
	\begin{tabular}{ccc} 
		\toprule[1.2pt]
		{Data} &  {Dev} & {Test} \\
		\hline
		{ 200h CS}  & 5.72 & 5.45 \\
		+ 500h CH & 4.37 & 4.28 \\
		+ 460h EN & 5.80 & 5.52 \\
		All & 4.33 &  4.32 \\
		\bottomrule[1.2pt]
	\end{tabular}
\end{table} 

\begin{table}[htb] 
	\centering 
	\caption{The MER/CER/WER (\%) of different transformer-based systems training with code-switching data. CH is the CER of the Chinese part and EN is the WER of the English part in the both dev and test data. PER refers to phoneme encoder layer. AEL refers to acoustic encoder layer.} \label{tab:aStrangeTable} %
	\renewcommand\tabcolsep{4.5pt}
	\begin{tabular}{ccccccc}
		\toprule[1.2pt]
		\multirow{2}*{model}& \multicolumn{3}{c}{Dev} & \multicolumn{3}{c}{Test}\\
		\cmidrule(r){2-4} \cmidrule(r){5-7}
		~ & All & CH & EN & All & CH & EN\\
		\hline
		Transformer & 12.52 & 10.20 & 31.21 & 11.63 & 9.48 & 29.30\\ 
		\hline
		Decoupled & \multirow{2}*{11.21} & \multirow{2}*{9.08} & \multirow{2}*{28.36} & \multirow{2}*{\textbf{10.51}} & \multirow{2}*{8.52} & \multirow{2}*{26.88}\\
		Transformer & ~ & ~ & ~ & ~ & ~ & ~ \\
		- PEL & 12.85 & 10.47 & 32.04 & 11.90 & 9.67 & 30.20\\	
		- AEL & 17.51 & 14.41 & 42.55 & 17.50 & 14.45 & 42.47\\
		\bottomrule[1.2pt]
	\end{tabular} 
\end{table}

\begin{table}[ht] 
	\centering 
	\caption{The MER/CER/WER (\%) of transformer with different training data.} \label{tab:aStrangeTable} %
	\renewcommand\tabcolsep{2.5pt}
	\begin{tabular}{ccccccccc} 
		\toprule[1.2pt]
		\multirow{2}*{model} &\multirow{2}*{Data}& \multicolumn{3}{c}{Dev} & \multicolumn{3}{c}{Test}\\
		\cmidrule(r){3-5} \cmidrule(r){6-8}
		~ & ~ & All & CH & EN & All & CH & EN\\
		\hline
		~ & 200h CS & 12.52 & 10.20 & 31.32 & 11.63 & 9.48 & 29.30 \\ 
		
		\multirow{2}*{Transformer} & + 500h CH & 12.17 & 9.64 & 32.55 & 11.42 & 8.99 & 31.37 \\
		
		~ & + 460h EN & 12.84 & 10.79 & 29.41 & 12.11 & 10.34 & 26.63 \\
		
		~ & All  & 11.94 & 9.70 & 30.02 & \textbf{11.21} & 9.15 & 28.10 \\
		\bottomrule[1.2pt]
	\end{tabular} 
\end{table}

\begin{table}[ht] 
	\centering 
	\caption{The MER/CER/WER (\%) of decoupled transformer with different pre-training data.} \label{tab:aStrangeTable} %
	\renewcommand\tabcolsep{2.5pt}
	\begin{tabular}{cccccccc} 
		\toprule[1.2pt]
		\multirow{2}*{model} & \multirow{2}*{Data}& \multicolumn{3}{c}{Dev} & \multicolumn{3}{c}{Test}\\
		\cmidrule(r){3-5} \cmidrule(r){6-8}
		~ & ~ & All & CH & EN & All & CH & EN\\
		\hline
		~ & 200h CS & 11.21 & 9.08 & 28.36 & 10.51 & 8.52 & 26.88 \\ 
		
		Decoupled & + 500h CH & 10.30 & 8.02 & 28.64 & 9.94 & 7.86 & 27.01 \\
		
		Transformer & + 460h EN & 10.99 & 9.22 & 25.31 & 10.42 & 8.64 & 24.97 \\
		
		 ~ & All & 10.21 & 8.24 & 26.08 & \textbf{9.63} & 7.76 & 25.00 \\

		~ & All + CS text & 10.10 & 8.13 & 26.00 & \textbf{9.52} & 7.72 & 24.25 \\
		\bottomrule[1.2pt]
	\end{tabular} 
\end{table}

\subsection{Results with Code-switching Data}
\label{ssec:subhead}
To verify the performance of the proposed model training with code-switching data. We compare the results of our method with the baseline model under the code-switching data. The results are shown in Table 2. The transformer's results show that the recognition performance on the Chinese part is better than English obviously. It is because the Chinese dominates the data set. And more Chinese training data is available for the model. We can find that performance of the proposed model has obvious improvement compared with the baseline. Our model achieves 9.63\% relative MER reduction compared with the baseline. And English WER and Chinese CER are improved by 8.26\% and 10.13\% respectively. This indicates that our method can model the distribution of code-switching speech better.

To examine how the acoustic and phoneme information affect the model performance, ablation experiments are conducted. As shown in Table 2, -PEL and -AEL refer to only pre-trained acoustic encoder and phoneme encoder is used in the proposed method respectively. The model -PEL has a similar performance as the transformer model. This means that the pra-trained encoder in our model can provide sufficient acoustic information. However, the model with only phoneme encoder can not achieve as good result as the baseline system. It shows that P2T model can not effectively deal with code-switching ASR tasks without acoustic information. The experimental results prove that it is beneficial to combine the two kinds of information.

\subsection{ Results with External Training Data}
\label{ssec:subhead}

Table 3 shows the results of the baseline model with external training data. It is clear that external monolingual data can reduce the error rate of the corresponding language in code-switching. However, the performance of another language is damaged to some extent. It is worth noting that external English training data even increase the overall MER of the model. The extra monolingual data negatively affect the model performance. The results demonstrate that monolingual data can not always improve the code-switching performance for the E2E model. When trained with all the data, the baseline achieves 3.61\% relative MER reduction compared with only code-switching training data.

To verify our model’s ability to use monolingual data, we pre-train the A2P network with different external data. The results are shown in Table 4. It is obvious that our method achieves greater improvement with external monolingual data than the baseline model. Although English data slightly increases the PER of A2P, the acoustic information learned from English monolingual corpus can still improve the overall performance of the model. For all data, our model achieves 8.37\% relative MER reduction compared with only code-switching training data.

To illustrate that the proposed model can effectively utilize unpaired code-switching text data, we use the code-switching training text and the corresponding  phoneme sequence to pre-train the P2T model. Then the whole model is fine-tuned with all data. The result shows that this strategy can also improve the performance of the model, albeit slightly. Overall, the proposed model provides up to 9.42\% relative reduction in MER compared with the only code-switching training data. And English WER and Chinese CER are improved 9.39\% and 9.78\% respectively. More code-switching text from data augment technology may further promote model performance.

\section{CONCLUSION AND FUTURE WORK}
\label{sec:majhead}

In this paper, we propose a decoupled transformer model to use monolingual paired data and unpaired text data to alleviate the problem of code-switching data shortage. In the model, multi-level attention mechanism is used to reduce the computational cost. The training and decoding process is as simple as the ordinary E2E model. The experimental results on code-switching data show that our method has consistent improvement compared with the baseline. It is an effective strategy for code-switching ASR tasks. In the future, we will pre-train the model with large amounts of unpaired text data, which from code-switching text generation technology.

\vfill\pagebreak

%\section{REFERENCES}
%\label{sec:refs}

\bibliographystyle{IEEEbib}
\bibliography{strings,refs}

\begin{thebibliography}{10}

\bibitem{muysken2000bilingual}
Pieter Muysken, Pieter~Cornelis Muysken, et~al.,
\newblock {\em Bilingual speech: A typology of code-mixing},
\newblock Cambridge University Press, 2000.

\bibitem{li2019towards}
Ke~Li, Jinyu Li, Guoli Ye, Rui Zhao, and Yifan Gong,
\newblock ``Towards code-switching asr for end-to-end ctc models,''
\newblock in {\em ICASSP 2019-2019 IEEE International Conference on Acoustics,
  Speech and Signal Processing (ICASSP)}. IEEE, 2019, pp. 6076--6080.

\bibitem{8682674}
B.~{Li}, Y.~{Zhang}, T.~{Sainath}, Y.~{Wu}, and W.~{Chan},
\newblock ``Bytes are all you need: End-to-end multilingual speech recognition
  and synthesis with bytes,''
\newblock in {\em ICASSP 2019 - 2019 IEEE International Conference on
  Acoustics, Speech and Signal Processing (ICASSP)}, 2019, pp. 5621--5625.

\bibitem{8462201}
S.~{Kim} and M.~L. {Seltzer},
\newblock ``Towards language-universal end-to-end speech recognition,''
\newblock in {\em 2018 IEEE International Conference on Acoustics, Speech and
  Signal Processing (ICASSP)}, 2018, pp. 4914--4918.

\bibitem{DBLP:conf/nips/VaswaniSPUJGKP17}
Ashish Vaswani, Noam Shazeer, Niki Parmar, Jakob Uszkoreit, Llion Jones,
  Aidan~N. Gomez, Lukasz Kaiser, and Illia Polosukhin,
\newblock ``Attention is all you need,''
\newblock in {\em Advances in Neural Information Processing Systems 30: Annual
  Conference on Neural Information Processing Systems 2017, 4-9 December 2017,
  Long Beach, CA, {USA}}, Isabelle Guyon, Ulrike von Luxburg, Samy Bengio,
  Hanna~M. Wallach, Rob Fergus, S.~V.~N. Vishwanathan, and Roman Garnett, Eds.,
  2017, pp. 5998--6008.

\bibitem{graves2006connectionist}
Alex Graves, Santiago Fern{\'a}ndez, Faustino Gomez, and J{\"u}rgen
  Schmidhuber,
\newblock ``Connectionist temporal classification: labelling unsegmented
  sequence data with recurrent neural networks,''
\newblock in {\em Proceedings of the 23rd international conference on Machine
  learning}. ACM, 2006, pp. 369--376.

\bibitem{graves2013speech}
Alex Graves, Abdel-rahman Mohamed, and Geoffrey Hinton,
\newblock ``Speech recognition with deep recurrent neural networks,''
\newblock in {\em 2013 IEEE international conference on acoustics, speech and
  signal processing}. IEEE, 2013, pp. 6645--6649.

\bibitem{chan2016listen}
William Chan, Navdeep Jaitly, Quoc Le, and Oriol Vinyals,
\newblock ``Listen, attend and spell: A neural network for large vocabulary
  conversational speech recognition,''
\newblock in {\em 2016 IEEE International Conference on Acoustics, Speech and
  Signal Processing (ICASSP)}. IEEE, 2016, pp. 4960--4964.

\bibitem{8462506}
L.~{Dong}, S.~{Xu}, and B.~{Xu},
\newblock ``Speech-transformer: A no-recurrence sequence-to-sequence model for
  speech recognition,''
\newblock in {\em 2018 IEEE International Conference on Acoustics, Speech and
  Signal Processing (ICASSP)}, 2018, pp. 5884--5888.

\bibitem{DBLP:journals/corr/abs-2006-00782}
Sanket Shah, Basil Abraham, Gurunath~Reddy M, Sunayana Sitaram, and Vikas
  Joshi,
\newblock ``Learning to recognize code-switched speech without forgetting
  monolingual speech recognition,''
\newblock {\em CoRR}, vol. abs/2006.00782, 2020.

\bibitem{DBLP:conf/interspeech/KhassanovXPZCNM19}
Yerbolat Khassanov, Haihua Xu, Van~Tung Pham, Zhiping Zeng, Eng~Siong Chng,
  Chongjia Ni, and Bin Ma,
\newblock ``Constrained output embeddings for end-to-end code-switching speech
  recognition with only monolingual data,''
\newblock in {\em Interspeech 2019, Graz, Austria, 15-19 September 2019},
  Gernot Kubin and Zdravko Kacic, Eds. 2019, pp. 2160--2164, {ISCA}.

\bibitem{DBLP:conf/interspeech/YilmazHL18}
Emre Yilmaz, Henk van~den Heuvel, and David~A. van Leeuwen,
\newblock ``Acoustic and textual data augmentation for improved {ASR} of
  code-switching speech,''
\newblock in {\em Interspeech 2018, Hyderabad, India, 2-6 September 2018},
  B.~Yegnanarayana, Ed. 2018, pp. 1933--1937, {ISCA}.

\bibitem{DBLP:conf/interspeech/ChangCL19}
Ching{-}Ting Chang, Shun{-}Po Chuang, and Hung{-}yi Lee,
\newblock ``Code-switching sentence generation by generative adversarial
  networks and its application to data augmentation,''
\newblock in {\em Interspeech 2019, Graz, Austria, 15-19 September 2019},
  Gernot Kubin and Zdravko Kacic, Eds. 2019, pp. 554--558, {ISCA}.

\bibitem{DBLP:conf/sltu/YilmazH18}
Emre Yilmaz and Henk van~den Heuvel,
\newblock ``Code-switching detection with data-augmented acoustic and language
  models,''
\newblock in {\em 6th Intl. Workshop on Spoken Language Technologies for
  Under-Resourced Languages, {SLTU} 2018, 29-31 August 2018, Gurugram, India},
  Shyam~S. Agrawal, Ed. 2018, pp. 127--131, {ISCA}.

\bibitem{DBLP:conf/interspeech/SriramJSC18}
Anuroop Sriram, Heewoo Jun, Sanjeev Satheesh, and Adam Coates,
\newblock ``Cold fusion: Training seq2seq models together with language
  models,''
\newblock in {\em Interspeech 2018, 19th Annual Conference of the International
  Speech Communication Association, Hyderabad, India, 2-6 September 2018},
  B.~Yegnanarayana, Ed. 2018, pp. 387--391, {ISCA}.

\bibitem{DBLP:conf/interspeech/ZhaoSRRBLP19}
Ding Zhao, Tara~N. Sainath, David Rybach, Pat Rondon, Deepti Bhatia, Bo~Li, and
  Ruoming Pang,
\newblock ``Shallow-fusion end-to-end contextual biasing,''
\newblock in {\em Interspeech 2019, 20th Annual Conference of the International
  Speech Communication Association, Graz, Austria, 15-19 September 2019},
  Gernot Kubin and Zdravko Kacic, Eds. 2019, pp. 1418--1422, {ISCA}.

\bibitem{DBLP:conf/interspeech/BaiYTTW19}
Ye~Bai, Jiangyan Yi, Jianhua Tao, Zhengkun Tian, and Zhengqi Wen,
\newblock ``Learn spelling from teachers: Transferring knowledge from language
  models to sequence-to-sequence speech recognition,''
\newblock in {\em Interspeech 2019, 20th Annual Conference of the International
  Speech Communication Association, Graz, Austria, 15-19 September 2019},
  Gernot Kubin and Zdravko Kacic, Eds. 2019, pp. 3795--3799, {ISCA}.

\bibitem{DBLP:conf/icassp/ChenLLY18}
Zhehuai Chen, Qi~Liu, Hao Li, and Kai Yu,
\newblock ``On modular training of neural acoustics-to-word model for
  {LVCSR},''
\newblock in {\em 2018 {IEEE} International Conference on Acoustics, Speech and
  Signal Processing, {ICASSP} 2018, Calgary, AB, Canada, April 15-20, 2018}.
  2018, pp. 4754--4758, {IEEE}.

\bibitem{DBLP:journals/corr/abs-2007-05916}
Xian Shi, Qiangze Feng, and Lei Xie,
\newblock ``The {ASRU} 2019 mandarin-english code-switching speech recognition
  challenge: Open datasets, tracks, methods and results,''
\newblock {\em CoRR}, vol. abs/2007.05916, 2020.

\bibitem{DBLP:conf/icassp/PanayotovCPK15}
Vassil Panayotov, Guoguo Chen, Daniel Povey, and Sanjeev Khudanpur,
\newblock ``Librispeech: An {ASR} corpus based on public domain audio books,''
\newblock in {\em 2015 {IEEE}, {ICASSP} 2015, South Brisbane, Queensland,
  Australia, April 19-24, 2015}. 2015, pp. 5206--5210, {IEEE}.

\bibitem{DBLP:conf/interspeech/ParkCZCZCL19}
Daniel~S. Park, William Chan, Yu~Zhang, Chung{-}Cheng Chiu, Barret Zoph,
  Ekin~D. Cubuk, and Quoc~V. Le,
\newblock ``Specaugment: {A} simple data augmentation method for automatic
  speech recognition,''
\newblock in {\em Interspeech 2019, 20th Annual Conference of the International
  Speech Communication Association, Graz, Austria, 15-19 September 2019},
  Gernot Kubin and Zdravko Kacic, Eds. 2019, pp. 2613--2617, {ISCA}.

\bibitem{DBLP:journals/corr/abs-1804-00015}
Shinji Watanabe, Takaaki Hori, Shigeki Karita, Tomoki Hayashi, Jiro Nishitoba,
  Yuya Unno, Nelson Enrique~Yalta Soplin, Jahn Heymann, Matthew Wiesner, Nanxin
  Chen, Adithya Renduchintala, and Tsubasa Ochiai,
\newblock ``Espnet: End-to-end speech processing toolkit,''
\newblock {\em CoRR}, vol. abs/1804.00015, 2018.

\end{thebibliography}

\end{document}